\documentclass[article,
showpacs,superscriptaddress,floatfix,preprint]{revtex4-1}

\usepackage[OT1]{fontenc}
\usepackage[usenames,dvipsnames]{color}
\usepackage[latin1]{inputenc}
\usepackage[english]{babel}
\usepackage{graphicx}
\usepackage{color}
\usepackage{siunitx}
\usepackage{hyperref}
\usepackage{xspace}
\usepackage{upgreek}
\usepackage{epstopdf}
\usepackage{amssymb,amsmath,verbatim,ulem}
\usepackage{braket}
\begin{document}

\title{Laser interferometry based on atomic coherence}

\author{Mangesh Bhattarai}
 \affiliation{Department of Physics, Indian Institute of Science, Bangalore - 560012}
 \author{Sumanta Khan}
 \affiliation{Department of Physics, Indian Institute of Science, Bangalore - 560012}
\author{Vasant Natarajan}
 \email{vasant@physics.iisc.ernet.in}
\affiliation{Department of Physics, Indian Institute of Science, Bangalore - 560012}
\author{Kanhaiya Pandey}
 \affiliation{Department of Physics, Indian Institute of Technology Guwahati, Guwahati - 781039}

\begin{abstract}
	
		We demonstrate laser interferometry based on phase difference between the two arms of the interferometer. The experiments are done with a Cs atomic vapor cell at room temperature and use atomic coherence. The interference can be tuned from constructive to destructive by tuning the relative phase between the two arms. It is similar to the Michelson interferometer, but differs in the important aspect of allowing interference when the polarizations in the two arms are orthogonal. This would be a novel method for interfering two independent lasers, which even can allow interfering two independent lasers of completely different wavelengths---such as of UV and IR---and also possibly phase lock them.

\end{abstract}

\maketitle

\section{Introduction}

Optical interferometry is at the heart of many sensitive measurements of the relative phase of two electromagnetic fields of similar frequency. It has variety of applications in industry and in fundamental research. Recently gravitational waves have been detected using LIGO and VIRGO project \cite{AAA2017}, which are based upon large scale Michelson interferometry. Optical interferometry techniques are based upon interference in the free space between two electromagnetic fields of the same polarization and of similar frequencies. However if the frequency difference of the two fields is very large (in THz), it is hard to compare their phases. Many applications that require phase coherence between the different optical beams have employed several techniques to lock the phases of two (or more) independent lasers \cite{LYK2007,SFB2013,WYN1999,SCL1994,MAS2008,STH2001,CHC2019}. For study of coherence effects in light-atom interaction two lasers differing in frequency by 10 GHz have been phase locked \cite{WYN1999,SCL1994,MAS2008}. This requires a photodetector with bandwidth several times of 10 GHz. Lasers with large wavelength difference have also been phase locked by interfering the higher (but different) harmonics at nearby frequency \cite{CDH2014}. Another sophisticated and costly method is to use a frequency comb to lock different lasers to the different frequency lines of the comb \cite{STH2001,NAZ2014,SWH2015,CHC2019}.

In this work we investigate an atomic based optical interferometry which is based upon electromagnetically induced transparency (EIT), as a proof of principle which will be capable of comparing and locking two optical fields having large wavelength difference. EIT for chain or branching system does not depend upon the phase difference of the control and probe lasers but only the robustness of it depends upon their phase stability \cite{PAN2013}. However in the loopy systems EIT depends upon the phase difference between probe and control lasers \cite{MKR1991,KMR1992,KOK1999,MFO2002,SNP2018}. The loopy system provides two paths of excitation which interfere with each other constructively or destructively depending upon the phase difference between control and probe lasers. In this work the two paths of excitation are made by two opposite polarized lasers hence interference is observed between two oppositly polarized fields in sharp contrast to the Michelson interferometer.

\section{Theory}
\subsection{A qualitative introduction}
\begin{figure}
	\includegraphics[width=0.7\textwidth]{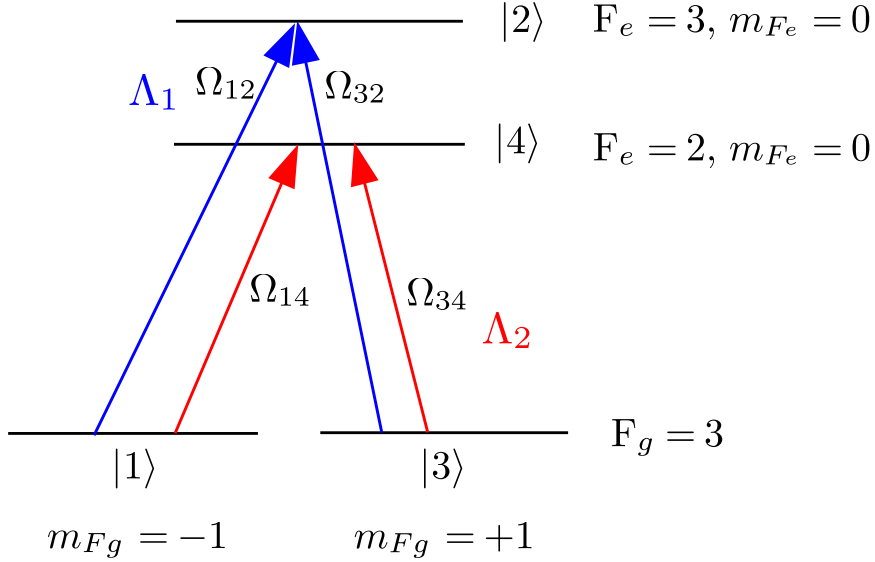}
	\caption{\label{fig:levels} A simplified representation of the energy levels involved in the system.}
\end{figure}

	Our system is composed of two $\Lambda$ systems sharing the common ground states as shown in Fig.~\ref{fig:levels}. The first subsystem $\Lambda_1$ is formed by two fields $\Omega_{12}$ and $\Omega_{32}$ that couple the ground states $\ket{1}$ and $\ket{3}$ to the excited state $\ket{2}$ . The second subsystem  $\Lambda_2$ has the states $\ket{1}$ and $\ket{3}$ coupled to $\ket{4}$ via the fields $\Omega_{14}$ and $\Omega_{34}$. We use coupling fields in $\Lambda_2$ stronger than those in $\Lambda_1$. In this setting $\Lambda_1$ will probe the influence of  $\Lambda_2$ in the atomic system. Because of stronger fields $\Lambda_2$ prepares the system in a non-absorbing state called a dark state $\ket{D_2} = \frac{1}{R_2} (\Omega_{34} \ket{1} - e^{i \alpha} \Omega_{14} \ket{3})$ \cite{KHA2019}, $\alpha$ is the phase difference between the two fields written explicitly and $R_2 = \sqrt{|\Omega_{14}|^2+|\Omega_{34}|^2}$. This phenomenon of preparation of system in a non-absorbing state is called coherent population trapping (CPT). Corresponding to the fields in $\Lambda_1$ the non-absorbing dark state  ($\ket{D_1}$) and the orthogonal absorbing bright  state ($\ket{B_1}$) can be written as,
\begin{equation}
\begin{aligned}
\ket{D_1} &= \frac{1}{R_1} (\Omega_{32} \ket{1} - e^{i \phi} \Omega_{12} \ket{3}) \\
\ket{B_1} &= \frac{1}{R_1} (\Omega_{12} e^{-i \phi} \ket{1} + \Omega_{32} \ket{3})
\end{aligned}
\end{equation}
where, $R_1 = \sqrt{|\Omega_{12}|^2+|\Omega_{32}|^2}$ and $\phi$ is the phase difference between $\Omega_{12}$ and $\Omega_{32}$.\\
The states $\ket{D_1}$ and $\ket{B_1}$ span the same Hilbert space as spanned by the ground states $\ket{1}$ and $\ket{3}$. Thus the state $\ket{D_2}$ can also be written as a linear combination of the dark ($\ket{D_1}$) and bright ($\ket{B_1}$) states of $\Lambda_1$ as 
\begin{equation}
\ket{D_2} = a \ket{D_1} + b \ket{B_1}
\end{equation}
The overlap of the atomic system in the state $\ket{D_2}$ with the bright state $\ket{B_1}$ given by the coefficient $b$ gives a measure of the absorption of the light fields in $\Lambda_1$. This can be calculated to be,
\begin{equation}
\begin{aligned}
|b|^2 &= |\Braket{B_1|D_2}|^2 \\
& = C_1+C_2 \cos{(\phi-\alpha)}
\label{eq:phase}
\end{aligned}
\end{equation}
where $C_1$ and $C_2$ are functions of the coupling Rabi frequencies. With $\alpha$ held as a constant (phase stable fields in $\Lambda_2$) the information of phase difference between the two fields in $\Lambda_1$ becomes visible in the absorption profile. This could be exploited to phase lock the two fields in $\Lambda_1$.

\subsection{Density matrix approach in bare state}
Now we provide details of the theoretical model using density matrix approach for probe laser absorption.
In the considered four-level-loopy system the electric field, corresponding to the transition $\ket{i}$ $\rightarrow$ $\ket{j}$ is $E_{ij}e^{i(\omega_{ij}t+\phi_{ij})}$, where $E_{ij}$, $\omega_{ij}$ and $\phi_{ij}$ are amplitude, frequency and the phase. 
Rabi frequency is defined as $\Omega_{ij} =d_{ij}E_{ij}e^{i\phi_{ij}}/\hbar$ for the transition $\ket{i}$ $\rightarrow$ $\ket{j}$ having the dipole moment matrix element $d_{ij}$. $\Omega_{ij}$ is a complex quantity which can be written as $|\Omega_{ij}|$ $e^{i\phi_{ij}}$, where $\phi_{ij}$ is the phase of the electric field associated with it. For dilute vapor cell the refractive index, $n$ for a probe laser $\Omega_{12}$ is related with the density matrix element, $\rho_{12}$ as $n=1+3\lambda_p^2N/(2\pi)(\Gamma_2/\Omega_{12})\rho_{12}$, where $\lambda_p$ (852~nm) is the wavelength of the probe laser and $N$ is atomic number density \cite{BLJ1995}. The absorption of the probe laser is represented by the imaginary part of $n$ and dispersion by the real part. We define the normalized absorption [$(\Gamma_2/\Omega_{12})$Im($\rho_{12})$] such that for stationary atoms, the absorption of a probe laser at resonance in the absence of all the control lasers  is 1.

The total Hamiltonian for this system is given as
\begin{equation}
\begin{aligned}
&H=\bigg[\sum^3_{i=1}\frac{\hbar\Omega_{i,i+1}}{2}\left(e^{i\omega_{i,i+1}t}+e^{-i\omega_{i,i+1}t}\right)\ket{i}\bra{i+1}\\ 
&+\frac{\hbar\Omega_{14}}{2}\left(e^{i\omega_{14}t}+e^{-i\omega_{14}t}\right)\ket{1}\bra{4}+h.c.\Big]+\sum_{j=1}^{4}\hbar\omega_j\ket{j}\bra{j}
\end{aligned}
\end{equation}
In the rotating frame with rotating wave approximation the above Hamiltonian will be,
\begin{align}
H&=\hbar\big[0\ket{1}\bra{1}-\delta_{12}\ket{2}\bra{2}-(\delta_{12}-\delta_{23})\ket{3}\bra{3}\nonumber\\
&-(\delta_{12}-\delta_{23}+\delta_{34})\ket{4}\bra{4}\big]\nonumber \\
&+\Bigg[\frac{\Omega_{12}}{2}\ket{1}\bra{2}+\frac{\Omega_{23}}{2}\ket{2}\bra{3}+\frac{\Omega_{34}}{2}\ket{3}\bra{4} \nonumber \\
&+\frac{\Omega_{14}}{2}e^{-i(\delta_{12}-\delta_{23}+\delta_{34}-\delta_{14})t}\ket{1}\bra{4}+h.c.\Bigg]
\label{Hami}
\end{align}

For general detunings of the lasers, the Hamilitonian $H$ is time dependent except for a particular condition when $\delta_{12}-\delta_{23}+\delta_{34}-\delta_{14}=0$. 

The time evolution of the density matrix, $\rho$ is given by Linblad master equation as 
\begin{equation}
\dot{\rho}=-\frac{i}{\hbar}[H, \rho]+L[\rho(t)] 
\label{Lin}
\end{equation}
where, $L[\rho(t)]$ is Linblad matrix and defined as below. 
\begin{equation}
\begin{bmatrix}
\Gamma_{21}\rho_{22}+\Gamma_{41}\rho_{44} &-\frac{\gamma^{dec}_{12}}{2}\rho_{12}&-\frac{\gamma^{dec}_{13}}{2}\rho_{13}&-\frac{\gamma^{dec}_{14}}{2}\rho_{14}\\
-\frac{\gamma^{dec}_{12}}{2}\rho_{21} &-\Gamma_{2}\rho_{22}&-\frac{\gamma^{dec}_{23}}{2}\rho_{23}&-\frac{\gamma^{dec}_{24}}{2}\rho_{24}\\
-\frac{\gamma^{dec}_{13}}{2}\rho_{31} &-\frac{\gamma^{dec}_{23}}{2}\rho_{32}&\Gamma_{23}\rho_{22}+\Gamma_{43}\rho_{44}&-\frac{\gamma^{dec}_{34}}{2}\rho_{34}\\
-\frac{\gamma^{dec}_{14}}{2}\rho_{41} &-\frac{\gamma^{dec}_{24}}{2}\rho_{42}&-\frac{\gamma^{dec}_{34}}{2}\rho_{43}&-\Gamma_{4}\rho_{44}\\
\end{bmatrix}
\label{Linterm}
\end{equation}
Where, $\Gamma_{ij}$ is the decay of the population from state $\ket{i}$ ($i=1,2$,.. to $4$) to state $\ket{j}$ ($j=1,2$,..to $4$) and $\Gamma_i$ is the total population decay rate of state $\ket{i}$. In the weak probe limit, the population dynamics between various levels can be ignored. In this case the crucial parameters are $\Gamma_i$ and $\Gamma_j$, i.e.the total decay rate of states, which also governs the decoherence rate ($\gamma^{dec}_{ij}$) between the two levels $\ket{i}$ and $\ket{j}$ as $\gamma^{dec}_{ij}=\frac{\Gamma_i+\Gamma_j}{2}$. In this particular case the value of $\gamma^{dec}_{12}=\gamma^{dec}_{14}=2\pi\times3$ MHz, which includes natural radiative decay of excited state, $\Gamma_2=\Gamma_4=2\pi\times 6$ MHz for Cs. 

From Eq. \eqref{Hami}, \eqref{Lin} and \eqref{Linterm} we have 16 coupled differential equations where $\rho_{ij}=\rho^*_{ji}$. We solve these coupled equation in steady state as done for multi-level systems \cite{PAN2013}.

We analyze the problem qualitatively by considering the effect of the coherence only between the levels and ignore the population transfer between them. So the time evolution of the population i.e. the diagonal terms of the density matrix such as $\rho_{11}$, $\rho_{22}$, $\rho_{33}$ and $\rho_{44}$  can be ignored with $\rho_{11}\approx1,\rho_{22}\approx\rho_{33}\approx\rho_{44}\approx0$. Similarly the time evolution of the off-diagonal terms $\rho_{ij}$ for $i=2; j=3,4$ and $i=3; j=4$ can be also ignored with $\rho_{23}\approx\rho_{24}\approx\rho_{34}\approx0$.\\
\begin{equation}
\begin{aligned}
\dot{\rho}_{12}&\approx i\frac{\Omega_{12}}{2}+i\frac{\Omega_{23}^*}{2}{\rho_{13}}-\gamma_{12}\rho_{12}\\
\dot{\rho}_{13}&\approx i\frac{\Omega_{23}}{2}{\rho_{12}}+i\frac{\Omega_{34}^*}{2}{\rho_{14}}-\gamma_{13}\rho_{13}\\
\dot{\rho}_{14}&\approx i\frac{\Omega_{14}}{2}e^{-i(\delta_{12}-\delta_{23}+\delta_{34}-\delta_{14})t}+i\frac{\Omega_{34}}{2}{\rho_{13}}-\gamma_{14}\rho_{14}\\
\end{aligned}
\end{equation}
Where, $\gamma_{12}=\left[\gamma^{dec}_{12}+i\delta_{12}\right]$, $\gamma_{13}=\left[\gamma^{dec}_{13}+i\left(\delta_{12}-\delta_{23}\right)\right]$,
\\ $\gamma_{14}=\left[\gamma^{dec}_{14}+i\left(\delta_{12}-\delta_{23}+\delta_{34}\right)\right]$, \\

In the four-photon resonance condition $\delta_{12}-\delta_{23}+\delta_{34}-\delta_{14}=0$, Hamiltonian \eqref{Hami} will be time independent. In order to satisfy the four-photon resonance condition for moving atoms, we choose laser $\Omega_{12}$ co-propagating to  $\Omega_{23}$ and $\Omega_{34}$ co-propagating to $\Omega_{14}$. In the steady state ($\dot{\rho}_{12}=\dot{\rho}_{13}=\dot{\rho}_{14}=0$)  we get,
\begin{equation}
\begin{aligned}
\rho_{12}&=\frac{i}{2}\frac{\Omega_{12}}{\gamma_{12}}+\frac{i}{2}\frac{\Omega_{23}^*}{\gamma_{12}}{\rho_{13}}\\
\rho_{13}&=\frac{i}{2}\frac{\Omega_{23}}{\gamma_{13}}{\rho_{12}}+\frac{i}{2}\frac{\Omega_{34}^*}{\gamma_{13}}{\rho_{14}}\\
\rho_{14}&=\frac{i}{2}\frac{\Omega_{14}}{\gamma_{14}}+\frac{i}{2}\frac{\Omega_{34}}{\gamma_{14}}{\rho_{13}}\\
\end{aligned}
\end{equation}
The above equation gives solution for $\rho_{12}$ as
\begin{equation}
\begin{aligned}
&\rho_{12}=\frac{\frac{i}{2}\frac{\Omega_{12}}{\gamma_{12}}}{1+\frac{\frac{1}{4}\frac{|\Omega_{23}|^2}{\gamma_{12}\gamma_{13}}}{1+\frac{1}{4}\frac{|\Omega_{34}|^2}{\gamma_{13}\gamma_{14}}}}-\frac{\frac{i}{8}\frac{\Omega_{23}^*\Omega_{34}^*\Omega_{14}}{\gamma_{12}\gamma_{13}\gamma_{14}}}{1+\frac{1}{4}\frac{|\Omega_{23}|^2}{\gamma_{12}\gamma_{13}}+\frac{1}{4}\frac{|\Omega_{34}|^2}{\gamma_{13}\gamma_{14}}}\\
\label{anasol}
\end{aligned}
\end{equation}
The first term $\rho_{12}$ in the above equation corresponds to the path of direct excitation $\ket{1}\rightarrow\ket{2}$ and further modified by the control lasers $\Omega_{23}$ and $\Omega_{34}$ which is known as EITA \cite {PAN2013}. The second term corresponds to the path of excitation to $\ket{2}$ through $\ket{1}\rightarrow\ket{4}\rightarrow\ket{3}\rightarrow\ket{2}$. These two paths interfere constructively or destructively depending upon the relative phase of these two. In this presented experimental configuration $\phi_{34}-\phi_{14}=\alpha$ is constant throughout the vapor cell (because of co-propagation) and is taken as 0 as the pump is x-polarized. Hence it cancels the phase factor for the $\Omega_{34}^*\Omega_{14}$ in Eq.~\eqref{anasol}. Hence the nature of the interference between two paths is finally governed by the relative phase of the two lasers $\Omega_{12}$ and $\Omega_{23}$ i.e. $\phi=\phi_{12}-\phi_{23}$.

In deriving analytical solution of the Eq.~\eqref{anasol} for $\rho_{12}$, we considered the no population transfer between the levels and considered only the effect of the coherence $\rho_{12}$, $\rho_{13}$ and $\rho_{14}$. In order to verify our assumption made in deriving the above formula we compare the complete numerical solution with the analytical solution and qualitatively we find a good agreement.  

\begin{figure}
	\includegraphics[width=0.75\textwidth]{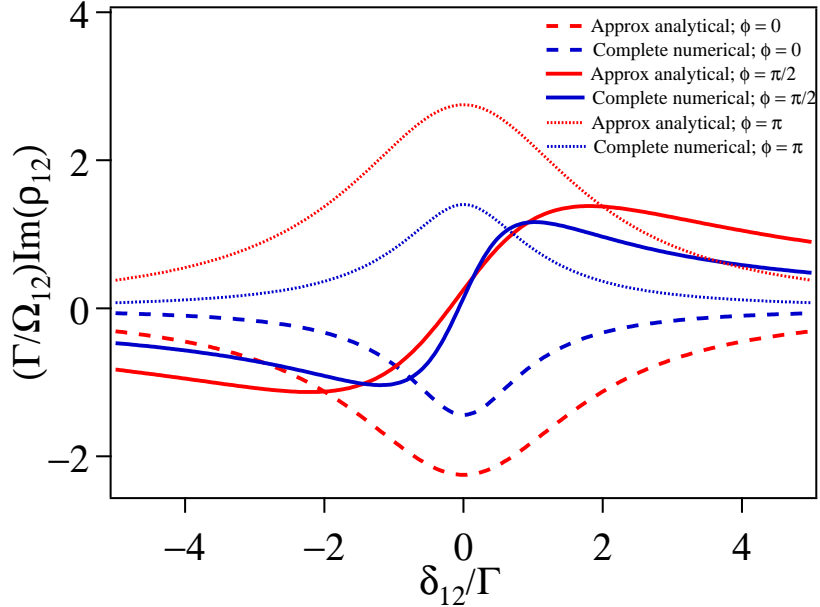}
	\caption{\label{fig:num_vs_ana} Comparison between complete numerical solution and approximated analytical solution (Approx analytical) of the normalized probe absorption ($\Gamma/\Omega_{12}$Im$\rho_{12}$) vs probe detuning $\delta_{12}$ with maintained four-photon resonance condition i.e. $\delta_{12}-\delta_{23}+\delta_{34}-\delta_{14}=0$ for three different phases, $\phi=0$, $\phi=\pi/2$ and $\phi=\pi$. The four-photon resonance condition is maintained by varying both $\delta_{12}$ and $\delta_{23}$ by equal amount and fixing $\delta_{34}=\delta_{14}=0$.}
\end{figure}    

\section{Experimental details}
\begin{figure}
	\includegraphics[width=0.7\textwidth]{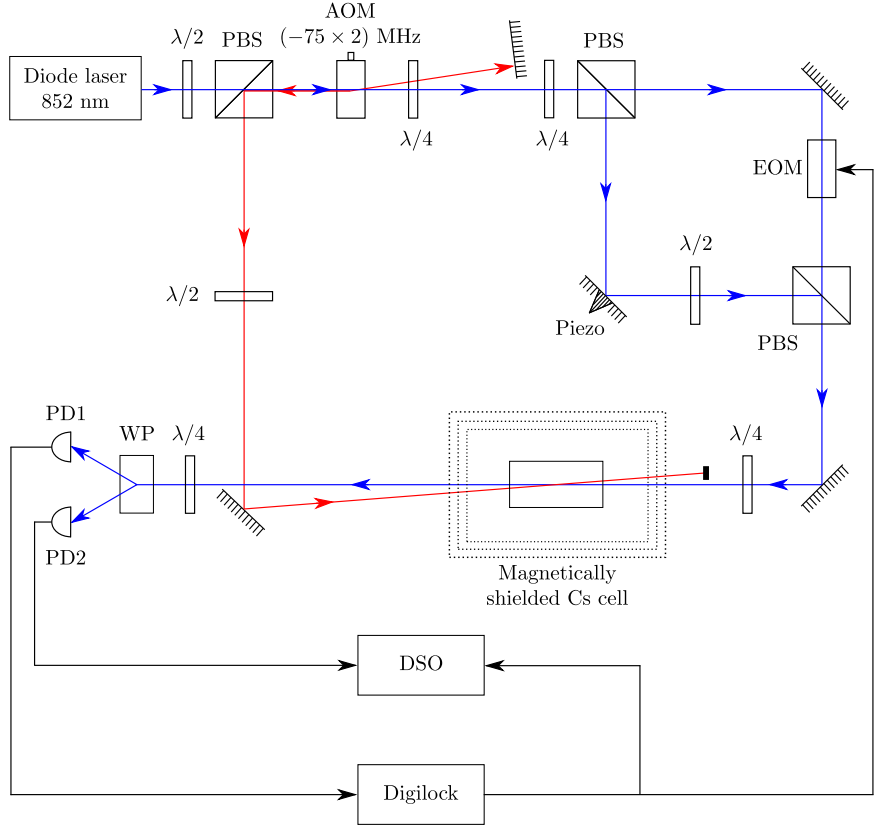}
	\caption{\label{fig:expt_setup} Schematic of the experimental set up. Figure key: AOM -- acousto-optic modulator; EOM -- electro-optic modulator; PBS -- polarizing beam splitter cube; WP -- Wollaston prism; PD -- photo-diode; $\lambda/2$ -- half wave retardation plate;  $\lambda/4$ -- quarter wave retardation plate. }
\end{figure}

Fig.~\ref{fig:expt_setup} gives an outline of the experimental setup. The light beam is derived from a Toptica DL Pro laser operating around 852 nm corresponding to the D$_2$ transition of $^{133}$Cs. It is locked to the $F_g = 3 \to F_e = 3$ transition in Cs using a saturated absorption spectroscopy setup. The beam from the laser is circular and has a 1/$e^2$ diameter of 3 mm. This beam is sent to an AOM which is used in a double-pass configuration to downshift the frequency by 151.232 MHz \cite{DAN2008}. This makes the double passed beam--- referred to as the pump beam (represented in red)--- resonant to the $F=3 \to F'=2$ transition. The un-shifted beam ---represented in blue in the Fig.~\ref{fig:expt_setup}--- goes into a Mach-Zehnder interferometer type setup. The beam is split into two orthogonally polarized beams using a polarizing beam splitter (PBS) cube. The p-polarized beam, henceforth referred to as probe beam, passes through an electro-optic modulator (EOM- eospace PM-0K5-10-PFA-PFA-850) along one arm of the interferometer. The EOM is fiber coupled with a PM fiber. The axis of the PM fiber and polarization of the coupling beam need to be aligned to prevent any amplitude modulation after polarizing optics. The s-polarized light on the other arm of the interferometer, which will be called control, after reflection from a piezo mounted mirror is recombined with the p-polarized probe beam on a PBS. The EOM and the piezo mounted mirror on the two interferometer paths can be used to independently change the phase difference between the two beams. A ramp signal from  a SRS DS345 function generator is used to drive the EOM to generate a linear phase shift. The piezo mounted on a mirror in the control beam arm is driven by a Thorlabs three axis piezo driver (MDT693) and any time varying signal is generated from a NI-PCI card. The co-propagating probe and control beams are converted to orthogonal circularly polarized light using a quarter wave retardation plate. A circular aperture of diameter 2.5 mm before the cell allows the most overlapped region of the beams to pass through. The typical power of probe or control used in the experiment is about 5-7 $\upmu$W. These beams are sent through a vapor cell containing Cs in vacuum. The vapor cell is cylindrical with 100 mm length and 25 mm diameter. It is housed within a three layer $\upmu$-metal shield to block the stray magnetic fields to $<$ 1 mG. Another quarter wave plate after the cell converts the circularly polarized light back to the corresponding linearly polarized components. The beams split into the p- and s- polarized components after a Wollaston prism which are then measured in the photo-diodes PD1 and PD2 (PDA36A-EC) respectively. The pump beam passes  into the cell counter-propagating with respect to the probe and control beams. These two beams intersect each other within the cell making a small angle of about 2$^\circ$. A half wave plate is introduced in the path of the pump beam before the cell. The wave plate can be used to introduce phase difference between the $\sigma^+$ and the $\sigma^-$ components of the beam.

\section{Results and discussion}
The probe and the control beam couple the $\sigma^+$ and $\sigma^-$ transitions between the sub-levels of $F_g=3$ and $F_e=3$. The linearly polarized pump couples the sub-levels of $F_g=3$ and $F_e=2$ forming the second lambda system. This transition is a cycling transition therefore the pump beam, usually higher in power, does not pump out atoms from the total system. All the transitions coupled by the beams form a complicated system but the essential physics could be entirely explained by considering the central lambda systems shown in Fig.~\ref{fig:levels}. 

\begin{figure}
	\includegraphics[width=0.5\textwidth]{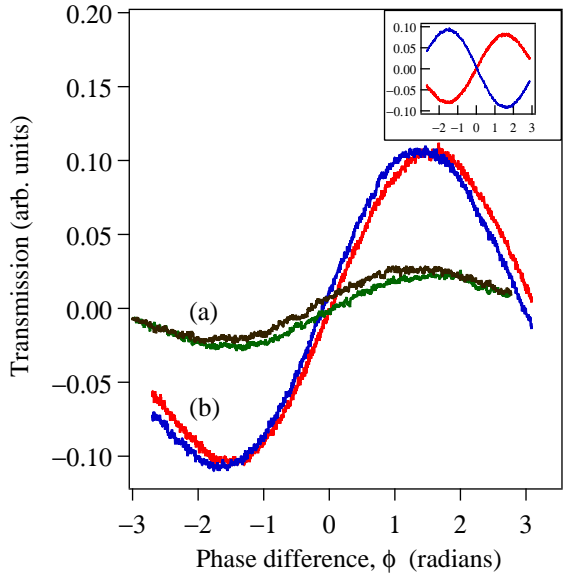}
	\caption{\label{fig:phaseVariation} The photo-diode signals recorded as a function of phase difference between the beam on the two arms of the Mach-Zehnder interferometer. The green and black trace (a) and the red and blue trace (b) are the two photo-diode outputs when the pump has power 100 $\upmu$W power and 1000 $\upmu$W respectively. Inset shows the photo-diode output when the laser is far off resonance. In all cases the two interferometer beams have equal power of 5 $\upmu$W.}
\end{figure}

The spectrum obtained as a function of phase difference due to EOM is shown in Fig.~\ref{fig:phaseVariation}. The voltage supplied to the EOM is converted to phase by fitting the PD1 output to a sinusoidal function and calibrating the separation between the maximum and minimum to be $\pi$ and setting the origin in the mid-point. The two pairs of curves (a) and (b) shown are for the counter-propagating pump beam at 100 and 1000 $\upmu$W respectively. We see that the two signals in each pair are in phase with respect to each other. It is reasonable as the two beams in a CPT configuration show the same absorption lineshape. The inset shows the case where the laser is far-detuned from the Doppler of the $3 \to F_e$ transition. This signal has no atomic origin but arises from the interference between the leaked portion of one polarization light in the other polarization arm. As expected from a Mach-Zehnder setup the two outputs are out of phase with respect to each other. The spectrum depicted in Fig.~\ref{fig:phaseVariation} is not very stable over time. Slow varying phase fluctuations introduced by environmental factors like flow of air, vibration of the platform and temperature and pressure variation on the EOM fiber makes the spectrum oscillate randomly. During data acquisition the spectrum was manually brought near the centre by changing the piezo voltage.

\begin{figure}
	\includegraphics[width=0.5\textwidth]{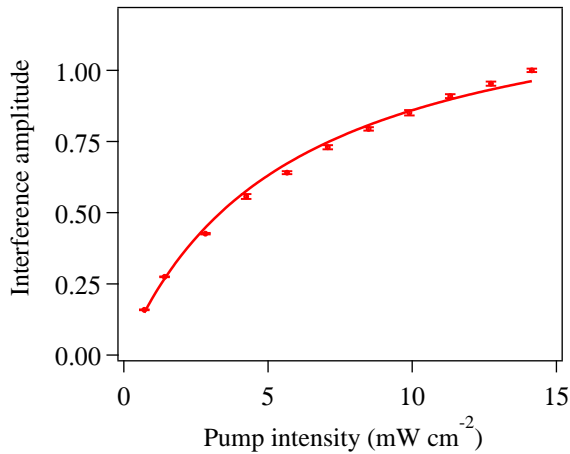}
	\caption{\label{fig:intensity_var} Normalized amplitude of the interference signal as a function of the pump intensity.}
\end{figure}

Variation of amplitude of the interference signal with the intensity of the pump beam is illustrated in Fig.~\ref{fig:intensity_var}. The open circles are the normalized amplitude of the interference recorded in PD1 (PD2 output shows the same behaviour) with probe and the control beam power of 7 $\upmu$W. The trend of the amplitude shows a faster increase near low pump  intensity and a slower increase as the intensities becomes higher. This could be understood in terms of the  scattering rate \cite{KKB17} of the pump beam which determines the number of atoms prepared in its dark states. Thus a higher intensity pump beam results in greater contrast in the interference. However when saturation effect become significant the the rise in interference amplitude slows down. The solid line in the figure is a fit to the scattering rate equation.\\

The spectrum in Fig.~\ref{fig:phaseVariation} also depends on the polarization of the pump beam. Polarization of a beam is a direct expression of the phase difference between its orthogonal circular components. When the phase difference between the two coupling beams in the $\Lambda_2$ system, represented by $\alpha$ in Eq.~\eqref{eq:phase}, is changed the spectrum shifts horizontally. This was observed experimentally as well (not shown here) using the half wave plate before pump. \\

\begin{figure}
	\includegraphics[width=0.5\textwidth]{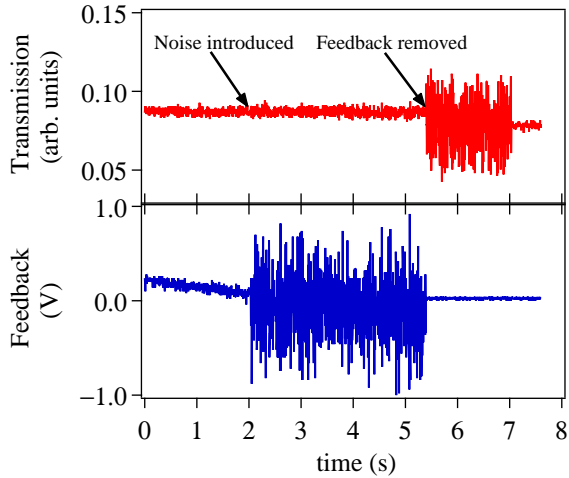}
	\caption{\label{fig:lockdemo}Demonstration of phase stabilization achieved via side-of-fringe locking with Digilock. Upper red trace is the noise in the interference pattern depending on whether the EOM is stabilized or not. Lower trace shows the random signal sent to EOM for active stabilization of the interference.}
\end{figure}

Next we attempted the phase locking of the probe and the control beam in this double lambda configuration. Locking was performed with the Digilock module from Toptica. The error signal for the phase stabilization was the spectrum similar to the one shown in the main plot in  Fig.~\ref{fig:phaseVariation}. This spectrum could be used to perform a side-of-fringe lock. Because both the spectrum are almost identical, we have used the output from PD1 as the input error signal into Diglock and monitored the output of PD2 to check the performance of the stabilization. The feedback for stabilization from Digilock is fed to the EOM which is also the channel that generates the scan. The probe and the control beam being derived from the same laser are perfectly coherent except for a slow varying phase fluctuation introduced by the environment. Thus to simulate a random phase relation between the two beams we send a Gaussian noise to the piezo generated via a NI PCI card at a sampling rate of 1 kS/s. We have verified that a noise on the piezo randomly shifts the spectrum of Fig.~\ref{fig:phaseVariation} horizontally. Fig.~\ref{fig:lockdemo} demonstrates the phase-locking achieved with the setup described. The upper trace shows the PD2 output and the lower trace is the feedback sent to the EOM from Digilock. The system was locked at a time before $t=0$ s. At around $t=2$ s , marked by the first arrow in the upper trace, a Gaussian noise is sent to the piezo as mentioned earlier. The feedback sent to the EOM itself is random and serves to completely remove noise from the interference pattern. In doing so the EOM has very accurately mapped the same phase noise onto the probe beam as in the control beam. Thus the two beams are perfectly phase coherent as their phase noise is perfectly correlated. After $t=5$ s the feedback sent to the EOM is removed while the noise is still sent to the piezo. Without an active stabilization the interference pattern now reflects the relative phase-noise between the beams as can be seen from the figure.

\begin{figure}
	\includegraphics[width=0.5\textwidth]{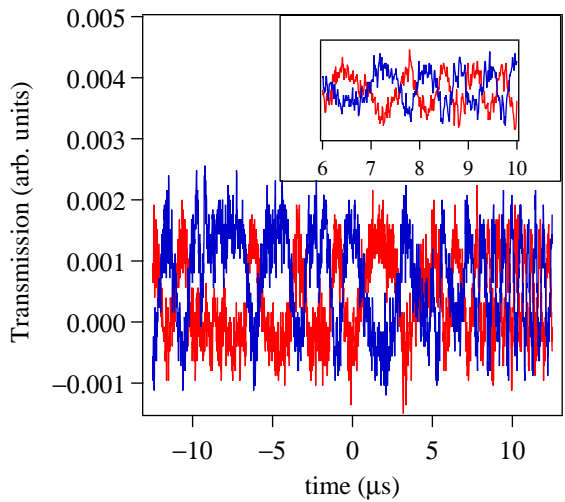}
	\caption{\label{fig:twoLasernoAtom} Interference signal without Cs cell between two independent lasers locked to the same $F=3 \to F'=3$ transition in Cs. The inset shows a zoomed view of the high frequency interference region.}
\end{figure}

\begin{figure}
	\includegraphics[width=0.5\textwidth]{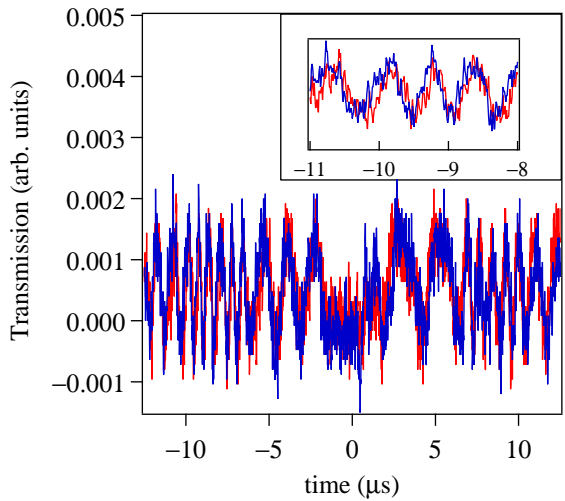}
	\caption{\label{fig:twoLaseryesAtom} Interference signal with Cs cell obtain between two independent lasers in the double Lambda configuration. The inset shows the region of high frequency interference magnified.}
\end{figure}

Till now all the beams were derived from a single laser. We now use the beam from a second independent laser and pass it through the EOM. This laser is locked to the $F=3\to F'=3$ transition in the D2 line of Cs. We observe the interference in two cases- without atom and with atom. The interference signals were observed with the highest bandwidth (10 MHz) setting of the photo-diodes. For the case of no atom interference signal is obtained due to the leakage of polarization as discussed previously. The fast varying interference signal reflects the phase noise between the two laser beams. As expected the interference signals obtained in the two channels PD1 and PD2 are out of phase-characteristic of a Mach Zehnder setup. This is shown in Fig.~\ref{fig:twoLasernoAtom}. The inset shows the magnification of a region of the spectrum. The interference pattern obtained with atoms and in the presence of the counter-propagating pump beam is illustrated in Fig.~\ref{fig:twoLaseryesAtom}. The inset of the figure showing the magnification of the high frequency region of the main graph clearly shows the in-phase behaviour of the two PD output signals which is characteristic of the double Lambda configuration. Besides this graph also gives information about the response time of the atom which is indeed quite vital when implementing phase locking. From the frequency of the interference signal we see that the atom exhibits a bandwidth larger than 2 MHz. In fact we have also observed that  with a step response given to the EOM (and where all beams were derived from the same laser) the interference signal showed a bandwidth of about 1.5 MHz for both cases of with and without atom (with PD operating with the specified bandwidth of 2.25 MHz). The same value of the bandwidth obtained without the atom just verified that the atomic system did not band-limit the signal. We can expect the bandwidth of the atom for this process to be of the order of the natural decay of the excited state because it takes some cycles of spontaneous emission for the formation of the dark state in CPT and EIT processes. Use of higher control power can further improve the bandwidth of the atomic system \cite{LIX1995,CDM1998}. \\

Though we have used degenerate sub-levels to form the legs of the $\Lambda$ systems in the experiment, it should be obvious that a non-degenerate choice would yield the same result.  This could be realized by considering that as long as the Raman detuning condition is satisfied CPT and EIT occur. In deriving the Eq.~\ref{eq:phase} we have not made any assumption about the degeneracy of the ground state. Another simple fact that supports this is that the form of the rotating wave approximated Hamiltonaian would be the same for degenerate and non-degenerate case, thus the physics should too. When non-degenerate ground states are used  the interference due to leaking polarization will occur at the beat frequency of the two beam whereas the interference due to our closed loop system would occur centered at zero frequency (DC). Here the atomic system itself does the demodulation of the phase noise. To implement such a system (non-degenerate ground states) the phase coherent lambda system which serves as the reference could be derived from a VCSEL or EOM --- for instance to use the two hyperfine levels of Cs --- to form one lambda and the two independent lasers to be phased locked form the other lambda. In addition the two beams of the arm of the Lambda gather different phase at different positions upon propagation ($(k_1-k_2) z$ is the phase difference at any point, $k_1$, $k_2$ are the two wavevectors and $z$ is the position along the direction of propagation). In this case if the beams of the two Lambda systems are co-propagating this accumulation of phase cancel as can be seen from Eq.~\eqref{eq:phase}.
\section{Conclusion}
We have demonstrated the phase dependence of absorption profiles in a closed loop system. This property can be used to phase stabilize two independent sources. Phase locking of sources with huge frequency difference that would otherwise be constrained by the bandwidth of the detector is no longer an issue as the phase noise would appear centered at 0 Hz. Though we have studied phase stabilization in just one type of closed loop system, several other systems of different topology could be implemented depending on the requirements.
\section{Acknowledgement}
K.P. acknowledges funding from SERB of grant no. ECR/2017/000781.

\end{document}